\documentstyle[pre,aps,preprint]{revtex}
\begin{document}
\title{Spreading of molecularly thin wetting films on solid interfaces}
\author{S.F.Burlatsky$^1$ \and  A.M.Cazabat$^2$ \and M.Moreau$^3$,
G.Oshanin$^4$ \and S.Villette$^2$}
\address{$^1$Department of Chemistry BG - 10,
 University of Washington,
 Seattle, WA 98195 USA; sfburlat@ringa.chem.washington.edu}
\address{
$^2$Laboratoire de Physique de la Mati\`ere Condens\'ee$^\diamond$,
Coll\`ege de France, 11
Place Marcelin Berthelot, 75231 Paris Cedex 05, France}
\address{
$^3$Laboratoire de Physique Th\'eorique des Liquides$^{\S}$,
Universit\'e Paris VI, 4 Place Jussieu, 75252 Paris Cedex 05,
France}
\address{
$^3$Centre de Recherche en Mod\'elisation Mol\'eculaire,
Service de Physique Statistique
 et Probabilit\'es, Universit\'e de
Mons-Hainaut, 20 Place du Parc, 7000 Mons, Belgium; gleb@gibbs.umh.ac.be}

\maketitle

\begin{center}
Abstract.
\end{center}
In this paper we study kinetics of spreading of thin liquid films on
solid interfaces. We present an overview of current experimental picture
and
discuss available theoretical approaches and their limitations.
We report some new experimental results on spreading of molecularly
thin liquid films and propose an analytically solvable microscopic
model, which reproduces experimentally observed behaviors and
provides a seemingly
plausible explanation of the underlying physical processes.

\pagebreak
\setcounter{page}{1}

\section{Introduction}

The spreading of liquid droplets on solid interfaces
 is  important
in many
technological and natural processes, such as lubrication,
painting, glueing, coating, emulsion, dyeing and oil recovery
from porous
rocks. In all instances, a precise knowledge
of conditions and laws
of spreading is required for efficient practical
applications.
Systematic experimental studies provide now a great wealth of
 information revealing
rich behavior dependent on the structural details and interactions
in the co-existing phases.
However,  currently available theoretical developments
don't explain all pertinent features of this
complex phenomenon, substantiated by experiments.
Particularly, still little is known
about
physical
mechanisms which cause growth of
molecularly thin films arising in the process of
spreading of  macroscopically large liquid droplets.

In this paper we focus on the particular issue of thin liquid
films spreading;  we discuss here results of earlier experimental studies,
as well as report some new ones,
which substantiate remarkable
universal (i.e. independent of the
liquid/solid system in question)
behavior of such
films and also propose an analytical description,
which reproduces experimentally observed
growth laws and
sheds the light on the underlying
physical processes.

The paper is outlined as follows:  Section II
contains
 a brief overview of
 current experimental and theoretical pictures of
wetting
films spreading.  In section III
we report results of recent ellipsometric experiments
on spreading of non-volatile liquid droplets,
 which produce
molecularly thin films.  We present  the data on the
time evolution of thickness profiles of spreading droplets
and also on the growth rate of the monolayer on
top of solid substrate.  In Section IV we formulate our analytical
model, write down basic equations and discuss their solutions
for both the case of spreading of  circular liquid droplets on solid
interfaces and growth of monolayers on vertical solid wall immersed
in liquid bath (capillary rise geometries). Finally, in Section
V we conclude with a summary of our results and discussion.

\section{An overview of experimental
 and analytical results}

When a liquid droplet is placed on a flat
 solid substrate,
three different interfaces come into play
 and three interfacial
tensions are involved; respectively, the
 tensions of the solid-liquid,
solid-vapour and liquid-vapour interfaces. The
 qualitative
behavior of the droplet is merely controlled
 by the sign of $S$
 - the so-called spreading parameter, which
equals
 the free energy difference between a bare solid
 and a solid covered by a thick
liquid layer. When $S > 0$ the droplet
 spreads spontaneously
and tends to shield the solid surface. Such a
situation is called the complete wetting. The
 case where $S$ is
negative is referred to as a partial wetting. Here
 the
liquid remains in the form of a droplet; it may contract
 or dilate but
 ultimately comes to an equilibrium bead-like shape and the liquid
droplet ceases to move.

The spreading of a pure, nonvolatile liquid droplet
 on a smooth,
homogeneous substrate is now well understood at
the macroscopic scales
\cite{pdga,amca,ball,legera,tan,broch,brochW,shana}.
 Continuum
hydrodynamic descriptions provide
general laws for the time evolution of such macroscopic
properties as the radius $R_{mle}$ of the macroscopic
 liquid edge,
 the height of the droplet, its
shape and the contact angle. These results suggest
 that in the
complete wetting case the time evolution
of the macroscopic
properties is rather insensitive to the specific
details
of the liquid/solid system in question and follows universal
time dependences. In particular,  $R_{mle}$
 was found to grow in proportion to $t^{1/10}$,
$t$ being time,
 in case of sufficiently small
 drops such that
the gravity effects are negligible; and a slightly  stronger dependence,
 $R_{mle} \sim t^{1/8}$, has been predicted and verified
experimentally for  situations in which gravity is
important.

The most recent progress in the complete wetting
case resulted
 from experimental works \cite{hesa,hesb,hesc}
 which
examined  kinetics of spreading on the
 mesoscopic
and microscopic scales. The salient feature here is
the appearance of a thin  film,  commonly referred to as the "precursor",
which extracts
from  the droplet and advances ahead of the macroscopic
 liquid edge. The thickness of the
film
  may vary considerably depending
on the particular liquid/solid system and may
 range from several
(molecular size) to hundreds of angstroms;
its  linear
extension is macroscopically
large (in the range of millimeters) at sufficiently large times.

Extensive ellipsometric studies \cite{hesa,hesb,hesc}
 (see also  recent works
 \cite{tib,vil,val,fra}
and pertinent
references therein)
have thoroughly examined the growth of the precursor
 film and
reached a rather surprising conclusion: as long as the
 droplet plays a role of a
reservoir for the film, the radius of the film $R(t)$
 obeys a universal law
$R(t) \sim \sqrt{t}$ regardless of the nature of
 the species involved.
To be more
specific, the same $\sqrt{t}$-law for growth of the precursor
 film
shows up in experiments involving
droplets of different
types of simple liquids, polymer or surfactant
 melts, and
carried out on
 different types of solid
substrates (either bare or grafted).
Furthermore,  it is observed also
in  the capillary rise
 geometries \cite{hesd,amcc},
in which a vertical solid
wall is put in contact with a bath of liquid.
In such experimental situation
 a film of microscopic thickness extracts from
 the macroscopic meniscus and creeps
upwards along the wall; the front of the film
 being planar compared to the
radially-symmetric one observed
for circular droplets deposited on a horizontal substrate.
The linear extension of the film again was found
 to grow in
proportion to $\sqrt{t}$ within a rather extended
time domain, until at very
 large times (which, in fact, may be several years)
and at large altitudes above the macroscopic meniscus
 the growth is truncated
 due to
gravity.
Finally, for several substrates an even
 more
remarkable behavior of spreading droplets was
 observed: experiments have shown that
 several
molecularly thin  films
may advance together, stacked on top of one
 another and thus
forming a stepped, "terraced"
shape of the drop  \cite{hesa,hesb,hesc} (see, e.g. Fig.1 of the present
paper).
Also  in this "terraced"
wetting case, at sufficiently short times the radii of
different monolayers were found
  \cite{hesa,hesb,hesc,val,fra} to
spread out as $\sqrt{t}$, with the
prefactor being a decreasing function
 of the distance from the substrate;
the bottom layer moves
outwards fastest, followed by the second
 from the bottom and etc.

Computer simulations are now able to reproduce
the  experimentally observed behavior
employing different types of simulation
 techniques. In
particular, MD simulations  with
 Lennard-Jones chain-like molecules,
 performed in \cite{jdcb,jdc,jdz},
exposed the spreading in the
form of distinct
layers with their radii growing as $\sqrt{t}$.
In the simulations in
 \cite{kon,luk} an Ising-type
 lattice-gas
model with Kawasaki dynamics has been  employed,
which has also
yielded
  the $\sqrt{t}$-law for growth of the
precursor film. Details of these and  earlier
numerical approaches
were recently reviewed in \cite{kas}.

Meanwhile several analytical approaches  are available, which
aim on the explanation of the
$\sqrt{t}$-law and of the "terraced" wetting phenomenon:\\
Joanny and
de Gennes \cite{joa} have developed a
continuum hydrodynamic theory
and found that the
$\sqrt{t}$-law originates from
a diffusive-type molecular motion
with  inhomogeneous
diffusion coefficient dependent on the
 local
disjoining
pressure.
This theory presumes, however, that the
film thickness
remains at least in the mesoscopic range,
where the continuum
hydrodynamics description
is still appropriate. Thus this approach does
not explain the growth of monolayers. Thin
liquid films on top of solid interfaces can not be
viewed as a true liquid phase. In such liquid/solid systems
the disordered liquid state contends with the
ordering potential
of the solid, which spans the liquid layer resulting in a markedly  different
behavior compared
to these occuring in the bulk liquids. For instance,
experiments reveal intriguing
effects of solid or
 glassy-like response to an external shear or anomalously high relaxation
times  \cite{grana,granb,isra,robb}.
 Cazabat et al.  \cite{amcc} proposed a phenomenological extension of
the hydrodynamic approach \cite{joa} for the
description of molecularly thin films spreading; in this description
basic equations of the theory by Joanny and
 de Gennes \cite{joa} have been adopted, but  a different
 origin
of frictional
forces has been assumed to account for the molecular behavior.\\
Further on, a  qualitatively different
 semicontinuum model, proposed
 by de Gennes
 and Cazabat \cite{pdgb},
treated
the drop as a layered structure, in which each
 layer is an
incompressible,
two-dimensional fluid. In this picture the motion
 of the fluid
molecules in the direction perpendicular to the
 layers is allowed
and the latter may grow by the accretion of
molecules at their edges
from the layers above
and below. Fluid particles experience an attractive
force from the
substrate
what makes the lower layers energetically more favorable
 and causes spreading
of the layers in the lower part of the drop.
This model predicts correct time behavior of the
advancing monolayers at long times, when one expects the
 difference between the
radii of neighboring layers to be large.  In this regime the radii
of monolayers are found to scale with time as $\sqrt{t/ln(t)}$.
Assumption of
incompressibility, however,  renders inadequate description
for short times, when the radii of monolayers are comparable.
Besides, the validity of the macroscopic
hydrodynamic description of dissipative forces, employed
 in the model by de Gennes
and Cazabat \cite{pdgb}, requires more extensive
microscopic justification. \\
Lastly, an interesting non-equilibrium statistical
 physics  description
 of the precursor
spreading
was based on the
solid-on-solid-model (SOSM) approximation \cite{abrc}.
Abraham et al \cite{abr,abrb}
have developed an interfacial model for the dynamics
 of a non-volatile
fluid edge, in which  the time evolution of the
liquid-vapour interface was analysed in terms of
 Langevin dynamics for the displacement of horizontal
 solid-on-solid strings at
increasing heights from the substrate. A free energy
 function associated with any
configuration of the interface  revealed
 a competition
 between surface
tension and substrate interaction. The model
allowed an analytical
solution, which showed an extraction of a precursor
 film and also "terraced"
forms
of  the
dynamical thickness profiles.
This approximation  predicted, however,
  a constant
velocity for the advancing precursor film,
i.e. $R(t) \propto t$, what  contradicts  apparently
 to experimental observations, and shows thus that  such an approach
discards some important aspects of spreading phenomenon.
To avoid
this inconsistency De Coninck et al. \cite{jdcc}
 elaborated a different,
 "columnar"  version of the SOSM, which takes
 into account entropic
repulsion effects,
but, nonetheless, obtained a precursor film
 extending linearly in time.

To summarize this section,
we
conclude that experimental studies
evidence the universal behavior of spreading monolayers;
 wetting films are found to grow  in proportion
to $\sqrt{t}$, regardless of the nature
of the  liquid/solid system.
The
origin and the underlying physics
of such a growth remain, however,
incompletely understood; computer simulations are now able to reproduce
it, but no convincing theoretical approaches have been, as yet,
developed.

\section{Experimental }

Experimental thickness profiles of spreading drops
 are conveniently studied using
spatially resolved ellipsometry. Ellipsometry
allows to measure
 the thickness of a film
on a substrate by analyzing the change of
 polarization of an optical
 beam at reflection.
For specific configurations, the optical beam
 can be focussed on the
 sample thus
improving the lateral resolution.

The present study was performed with a
polarization modulated
ellipsometer working at a
single wavelength ($6328$ $\AA$, He-Ne laser)
 and at Brewster angle.
 The substrates used were  oxidized
silicon wafers, either bare or bearing a grafted
 hydrophobic layer.
 The liquids involved were either light
polymers (polydimethylsiloxanes, trimethy-terminated,
abbreviated as PDMS)
 or nonvolatile
silane derivative, like tetrakis(2-ethyl-hexoxy)silane,
 abbreviated as TK. Specifically, the PDMS is a chain-like molecule,
 whose
molecular mass in our case is $9300$ (it contains $126$ monomers,
polydispersity
 index being $1.09$)
and the transverse size is $7$ $\AA$. The TK molecule is spherical,
 with a diameter around
$10$ $\AA$.
 The optical
index of the silica, of the grafted layer if any, and of
 the  liquids under study are close,
around
$1.4$, while the index of the underlying silicon is
 around $3.8$ for the red light.
 This high
contrast between silicon and the layers on top of it
 ensures a good thickness
resolution,
$0.1$ $\AA$ per measurement for a measurement time $20$ ms.

The lateral resolution of the setup is $25$ $\mu$m.
 At this scale, the
 substrate
roughness (typically $5$ $\AA$ rms, with a
characteristic length along
the substrate
$\approx 200 \AA$) is smoothed out and thus the
 thickness profiles are
not noisy. Note
that for the low thickness considered, an
independent determination of
the thickness
and of the optical index is not possible: All the
 layers on top of silicon
are seen as
a whole and the corresponding optical path is
converted into "thickness"
 by dividing by
the average bulk index value $1.4$ \cite{beag}.
The response of the setup being linear in this
range,
the relative error on the thickness is the relative
 error on the index. The baseline on the
profiles is the thickness of the silica layer.

Now,  in experiments we have monitored the time evolution
of two different properties -  the thickness profiles of spreading
droplets
and the radius of the first layer on top of solid substrate.
Results of experiments are plotted in Figs.1-4;  two
first figures concern the spreading of the TK droplets
at different relative humidities, while Figs.3 and 4
represent analogous data for the droplets composed of the PDMS
molecules.  Being interested mainly in the microscopic details
of the thickness profiles,  we have plotted the thickness profiles of
spreading droplets
(the $Y$-axis)
in the range of angstroms, while the radial size of the droplet
is presented on the millimeter scale, which causes a huge
disproportion between the scales on the $X$ and the $Y$-axis.
Consequently, the thickness profiles recorded at relatively short
times after the deposition of the droplet on the substrate are
out of scale.

In  Fig.1 we present the ellipsometric
profiles of the TK droplets spreading on a bare
silicon wafer  at relative
humidity $60$ $\%$.  The profiles were recorded consequently
at $30$mn, $5$h$45$,
 $22$h, $30$h, $4$
days, $5$ days and, eventually, $11$ days
 after the deposition of the drop.
It is clearly seen here that with increasing times
the profiles widen out and flatten, the
 volume of the drop being constant,
 forming well-defined
terraces of the step thickness $\approx 10$ $\AA$, i.e. comparable
 to the molecular size.
The baseline
 has been substracted,
i.e. zero thickness corresponds to the
surface of the silica.

In Fig.2 we plot experimental  thickness profiles
(top) and the time dependence of
the radius of the first layer (bottom) corresponding to the
 short time regime in
spreading of a relatively large droplet of TK (at low relative humidity, $20$
$\%$).
Profiles are measured at $1$h, $2$h$15$, $4$h, $8$h and $20$h$15$ respectively
after the deposition of the droplet. In contrast to the situation depicted in
Fig.1,
in this case the centre of the drop plays as a reservoir for the film during
the
whole period of time. We again stress the huge disproportion between the scales
on the $X$ and the $Y$-axis, to avoid the impression that the thicknesses and
the
 linear extensions
of layers are comparable to each other.
The baseline has not been substracted, i.e. the thickness of the
silica layer is $\approx 19$ $\AA$.  On bottom of the Fig.2 we plot the
 radius $R(t)$ of the
first layer versus the square root of the elapsed time. Straight line,

\[R(t) = R_{mle} \; + \;
\sqrt{\alpha t}, \quad (1)\]
with
$\alpha \approx 1.2 \times 10^{-10} m^{2}s^{-1}$ gives the fit
of experimental data.

Fig.3 describes the thickness profiles and the time dependence of the precursor
film
radius in case of the PDMS droplet spreading on a grafted hydrophobic surface;
the silica is now
covered by a layer of trimethyl groups. The total thickness of silica
 and grafted layer is $\approx 23$ $\AA$.  The thickness of the layer
 on the substrate is
$\approx 7$ $\AA$, which is the transverse size of the PDMS molecule;
this means that the
PDMS chains are lying flat on the top of the substrate. On bottom
 of Fig.3 we plot $R(t)$
 versus the square root of time. The straight line gives
the fit of experimental results by the  function in Eq.(1),
in which $\alpha \approx 5.5 \times 10^{-11} m^{2}s^{-1}$. The last
experimental
point
is off the straight line, because the "reservoir" has disappeared to this
moment.
 The radius
will now stay constant for months.

In Fig.4 we present experimental thickness profiles obtained during spreading
of
the PDMS
droplets in the conditions, in which the bare silica is exposed to high
relative
humidity
(RH = $90$ $\%$) and therefore is covered by a thin adsorbed layer of water.
Here, the
thickness
of water is $\approx 4$ $\AA$, i.e. of the order  of
 the transverse size of the PDMS
molecules.  The total thickness of silica and water
 is $\approx 18$ $\AA$. In this case
a single monolayer of a constant thickness shows up,
 advancing ahead of the macroscopic
liquid edge.  The macroscopic edge does not move significantly
 during the whole elapsed time.
The curves
on the
top of Fig.4 correspond to the profiles of a large droplet
at $1$h$30$, $2$h$30$ and $5$h after deposition; the curves on the bottom of
Fig.4
describe the thickness profiles of a small drop after
 $2$h, $3$h$30$ and $5$h$15$ (monolayer).
The time dependence  of the radius of the monolayer is fitted by Eq.(1)
 with $\alpha \approx 10^{-9} m^{2}s^{-1}$, which is
substantially greater  than that obtained in the previous case
(Fig.3).  The reason
is that the friction between spreading molecules and the surface is here
considerably
decreased, without changing the interaction in a significant way \cite{vil}.

\section{Analytical description of spreading of monolayers
on solid interface.}

In this section we formulate a microscopic, analytically solvable model
describing  spreading of molecularly thin
wetting films, write down basic equations and discuss their
solutions for both  the case of capillary rise geometries and
circular droplets deposited on a horizontal substrate. Some aspects
of this model and of the solutions have been  discussed
 in \cite{burc,burd}.

\subsection{The model}

The analytical model includes three basic points, which can
 be succinctly formulated as follows:

First,  we suppose that
the wetting film is in equilibrium
 with the bulk
liquid (the macroscopic part of the droplet deposited
 on a horizontal substrate
 or the liquid bath in case of capillary rise geometries),
which acts as a reservoir of particles and feeds the film.
We stipulate that the "reservoir"
has an infinite capacity  and thus the bulk liquid
maintains a constant density of fluid particles at the
macroscopic liquid edge (see Fig.5).

Second, dynamics
of fluid particles on top of solid interface is viewed
as a symmetric, activated random
hopping motion, constrained by hard-core interactions.
In such a picture the wetting film
can be thought off as being a hard-sphere fluid adsorbed on solid interface;
the attractive interactions between
the fluid particles are, as yet, discarded.

Third, we account for these cohesive interactions
by introducing an impenetrable "liquid-vapour" interface,
which encloses
the hard-sphere
fluid and influences dynamics of fluid particles
being
in the vicinity of the interface. The physical
 properties of
this "liquid-vapour"
interface are described in
terms of the SOSM approximation \cite{abr,abrb}; using some
 physical arguments we
relate the surface tension of this interface to the overall
 cohesion energy of the
particles in the film.

Now, let us describe the model more precisely:

\begin{itemize}
\item Fluid density at the macroscopic
 liquid edge
\end{itemize}
We denote the
 density of fluid
particles at  the macroscopic
 liquid edge (MLE) (see Fig.5) as  $\rho_{0}$.
The value of  $\rho_{0}$ can be estimated
 employing essentially the same type of
 reasonings as
those used for the derivation of the  Langmuir
 adsorption isotherm
\cite{avey}:
Suppose that one has  a "vacancy" directly at
 the macroscopic liquid edge
 and a fluid
particle in the bulk phase,  being at the
 distance of a single "jump"
from the MLE.
Let us denote now
as  $E_{\downarrow}$ the energy gained by
  moving this particle
onto the vacancy.
Then, in the limit $\beta E_{\downarrow} \gg 1$,
$\beta = 1/ k T$, we find  that $\rho_{0}$ is simply
\[
\rho_{0} \; \approx \; 1 \; -
\; \exp(- \beta E_{\downarrow}) \quad (2)
\]
In what follows we don't specify the value of $E_{\downarrow}$, which
depends on the particular details of liquid/solid system in question.
We merely note that it is dependent on both the
 liquid-liquid and liquid-solid interactions
and is determined by two different factors.
The positive contribution to $E_{\downarrow}$
is due to
the presence
 of attractive
interactions between the fluid particles
and the atoms of the solid  - a jump of a fluid particle downwards to the
solid interface decreases energy.
The
second, negative
factor stems from the presence of cohesive
 liquid-liquid interactions;
it equals the
energy loss due to  breaking
cohesive  bonds with several fluid
molecules,
since for the particles
being in the bulk phase
 the number of neighbors is greater than that
 for particles
directly on the
solid.  In case of chain molecules, this factor
includes the contribution associated with the work
 required to detach
the macromolecule from the melt of intertwined polymers.

\begin{itemize}
\item Dynamics of the film
 particles on the solid interface
\end{itemize}
We employ the conventional
 picture of such dynamics (see e.g. \cite{avey,blake}
 and references therein) and
view the motion of particles  as an activated
random hopping transport, constrained
by hard-core interactions, between
the local minima
of a wafer-like array of
potential wells (wavy line in Fig.5).
The reason why such wells  occur may be twofold:
  On one hand, the film's particles
move on the solid interface and thus experience
 the ordering potential
of solid atoms. On the other hand,
such wells arise because of the
 mutual, collective interactions of particles
 in the film
(as for the motion in bulk liquids).
Without going into details of particle-particle
 and particle-substrate
interactions, we suppose that for the
transition to  one of neighboring potential
 wells a particle
has to overcome a potential
barrier. This barrier does not create
 a preferential hopping
 direction, but
results in a finite time interval $\tau$ between
 the consecutive
hops, defined through the Arrhenius formula.  The
interwell distance (denoted as  $a$) may
be related either to the radius of the
 repulsive part of the
 particle-particle
interactions or to the spacing between
 the atoms of the substrate.

We assume that all particles (except the particles
 at the edge of the film, which  are
in immediate contact with the "liquid-vapour" interface;
 we will call these - the "boundary
particles")
have symmetric transition
rates: for these a  probability of hop in
 any of four directions is $1/4$.
Then, the
diffusion coefficient of particles on solid
  is defined through the
 parameters $a$ and
$\tau$ as $D = a^{2}/4 \tau$, where both $a$ and
$\tau$ depend on the form of liquid-liquid
and liquid-solid interactions.
Otherwise stated,
$D = k T/\gamma$, where
$k$ and $T$ are the Boltzmann constant
 and the temperature respectively,
while $\gamma$ is  the effective
friction coefficient for motion in a
 liquid monolayer on the solid interface.
Hard-core interactions
constrain the particle hopping motion;
no two
particles can simultaneously occupy
the same well.
Thus a hop onto a well, already occupied by another
fluid particle
is forbidden.

\begin{itemize}
\item Dynamics of the boundary particles
\end{itemize}

Now we define dynamics of the boundary particles (BP),
which will in some
aspects
 be different from that of
particles in the film.
First, for the BP the hops in the direction to the MLE
are constrained by the hard-core interactions,
 while  hops away from the MLE are
always  unconstrained. Second,  the boundary particles
 move in the vicinity
of an effective
 "liquid-vapour" interface which influences their motion
 exerting
a constant "restoring" force directed towards the MLE.
We thus
stipulate that for the BP
the hops which increase the distance from the MLE
occur with smaller probability ($p$)
than  hops which decrease this distance (such
 hops occur with the probability
$q$, $p < q$).  The values of $q$ and $p$ are
 constant in time and are independent
of the radius of the film.

Adopting the SOSM approximation \cite{abr,abrb,jdcc}, we may
 readily establish the relation between
the magnitude of this "restoring" force (and, consequently,
between the ratio $p/q$), and  the surface tension of the interface.
In this approximation, the energetic "cost"  $F(R(t),h_{1},h_{2}, ... , h_{N})$
of a particular configuration of an interface
(discretized as shown in Fig.5) with a fixed set of the layers radii,
 $R(t)$ and $\{h_{j}\}$, is
given by

\[F(R(t),h_{2},h_{3}, ... , h_{N}) \; = \;  P(|R(t) - h_{1}|) \; + \; \sum_{j =
1}^{N - 1}
 P(|h_{j+1} - h_{j}|), \quad (3)\]
where $P(x)$ describes the interaction energy between the neighboring layers;
$P(x)  \approx  J_{1} x^{2}$ for sufficiently small $x$ (elastic interface)
and $P(x) \approx
J |x|$ for $x$ being large \cite{abr,abrb,jdcc}. The parameter $J$ is
proportional
 to the surface tension.
Suppose next the situation as depicted in Fig.5, in which a single monolayer
of length $R(t)$ appears, such that
$R(t) \gg h_{1}$; $h_{1} \approx h_{j} \approx R_{mle}$.
Then, it follows from Eq.(3) that
the cost of surface energy for having a precursor film of radius $R(t)$ is
simply

\[F(R(t),h_{2},h_{3}, ... , h_{N}) \; \approx \; J R(t), \quad (4)\]
i.e. it increases linearly with $R(t)$. This means, in turn,
 that the interface exerts
 a $constant$,
 independent
of the film radius,  pressure on
 the film directed towards the
MLE.  In other words,
the BP experience an action of a
constant force $f$,
$f = - \; \partial  F/\partial R(t) = -  \; J$,
which favors its hops in the direction to
 the MLE. Let us stress that in
this picture
only the boundary particle is subject
 to a surface-induced force;
all other particles in the film
don't "feel" the presence of the
interface and thus have symmetric
transition rates.

Let us discuss now the physical meaning of the
"surface tension" parameter $J$.
The microscopic origin of the
asymmetric hopping rates
stems from the  mutual interactions
 between the particles in the film.
Typical interactions
in real systems are characterized
 by a harsh repulsion of a
 hard-core type
at short scales and attraction at
 longer distances.
Now, the  hops of
 the BP away from the MLE and the hops in the direction of
the MLE do not
change the number of particles at
 a given $Y$ but result in
stretching or shrinking of
the film.
Thus the change in the length of
 film comprising a fixed number of
particles results in the change
 of energy.
Stretching of the film
will lead to an increase of energy.
 Conversely, shrinking of the film
decreases the
interparticle distances  and thus
 results in a decrease of energy.
In other words,  the presence of
the  particle-particle
attraction results in correlations
 between the local transition rates and
spatial distribution of particles
- - these tend
to move  towards the spatial regions
 in which the particle
 density is
high. Since the density
is maximal at the MLE and decreases
 with an increase of distance,
the particles in  the film
experience, on average,  an action
 of a force which is directed to the MLE.
In our model  this circumstance is
taken into account in a mean-field fashion
by introducing an integral (over all
 particles of the film)
 force which acts on
the BP only and which is equivalent
 to the presence of a SOSM-type
 interface with some
effective surface energy.  To avoid
confusion with the surface tension $J$, which
appears in Eqs.(3) and (4), we will denote this
effective surface energy as $W_{\leftarrow}$.
In view of previous
 discussion, $W_{\leftarrow}$, which
 is the difference of the energies
gained and lost
due to the hop of the BP away from and
in the direction to the MLE, will be defined as
 the
 work required
to transport a vacancy from the
 edge of the film to the MLE. In contrast to the parameter
$E_{\downarrow}$, the effective surface energy $ W_{\leftarrow}$
is thus dependent only on the liquid-liquid interactions.
Using
detailed balance arguments we get the following
relation between $p$, $q$ and $W_{\leftarrow}$,
\[\frac{p}{q} \; = \;
\exp( - \beta W_{\leftarrow}) \quad (5)\]
We note, finally, that by definition
 $W_{\leftarrow}$  equals the
 difference
 of the
potential energy of vacancy placed at
 the MLE and the potential
energy of vacancy at the edge of the
 film,
and hence  is independent of the radius
 and the mass of the film,
 provided that
 $R(t) \gg a$.

\subsection{Basic equations.}

We turn now to the mathematical description
 of the  film growth focusing first on the case of
{\it capillary rise geometries}.

To specify positions of the wells, we introduce a pair of
perpendicular coordinate axis, in which the $Y$-axis will
be parallel to the MLE, i.e. will define the horizontal
 position of a given well,
while the $X$-axis will measure
the altitude of a given well above the MLE.  Here $R(t)$ gives the
height of the film relative to the origin; $R_{mle}$ (which is
a horizontal
straight line) separates the film and the macroscopic meniscus and
 the difference $R(t) - R_{mle}$
defines the linear extension of the film above  the macroscopic
meniscus.
Further on,
we define the variable
$\eta(X,Y,t)$ -  the time-dependent
 occupation variable of the
well with coordinates $X$ and $Y$. This variable may assume
two possible values;
it equals $1$ if the well
 is occupied and $0$ if
the well is empty.

We note now that the $X$ and the $Y$-dependences of
 $\eta(X,Y,t)$  are distinctly different. Along the $X$-axis
we have
 a
reservoir of particles, which maintains
fixed occupation of all
wells  with $X =  R_{mle}$, and
well-defined constant "restoring" force acting on the
 BPs.
Consequently, we may expect that the
 $X$-dependence of
$\eta(X,Y,t)$ is regular.
In contrast, the $Y$-dependence may  stem only out of
fluctuation effects;
the uniform boundary at the MLE insures
 that there is no regular
dependence
on the $Y$
coordinate and only
the particle dynamics may cause
fluctuations in $\eta(X,Y,t)$ along
the $Y$-axis.
In our present analysis we will disregard
 these fluctuations
 and suppose that
the occupation variable varies along
 the $X$-axis only, i.e.
$\eta(X,Y,t) = \eta(X,t)$, and the film front
is  a horizontal straight line.
We stress that an assumption of such a type
is quite consistent with
 experimental observations
\cite{hesa,hesb,hesc},
which show
that for sufficiently smooth
substrates and liquids
 with low volatility the
width of the film front is very narrow.
We also remark that regular $Y$-dependence
may arise in the situation under study, if one applies,
for instance, an upwardly directed temperature gradient
along the $X$-axis \cite{anne}. In this case of a "forced spreading",
the film's front undergoes a fingering instability resulting in a
nearly periodical $Y$-dependence \cite{anne}.  Appearance of such
effects, in general, requires very special conditions \cite{legera,anne},
which are assumed to be absent.

In neglect of the fluctuations
along the $Y$-axis  we are faced to consider an effectively
 one-dimensional
 problem in which
 the presence of the $Y$-direction will be
 accounted only
through the particles' dynamics.
Then, $\eta(X,t)$ can be viewed as
 a local time-dependent
 variable describing occupation of the
 site $X$ in a
$stochastic$ $process$ in which
 hard-core particles perform
 hopping motion (with the
time interval $\tau$ between the
consecutive hops) on a one-dimensional
lattice of spacing $a$ (see Fig.5).
All particles, except
 the BP, have
 probabilities
 $1/4$ for
hops from $X$ to $X \pm a$, and
 probability $1/2$ to stay at  $X$
(arising from the motion along the $Y$-axis).
The BP, being
 at  $X$, may jump to $X + a$
with
probability $p$ and to $X - a$
 with probability $q$, provided
 that this site is
vacant; and may remain  at $X$ with
probability $1/2$. Further on,  a source at
 $X = R_{mle}$
maintains a
fixed occupation of this site.
This process is a generalization of a
"directed walk in a
lattice gas" model,
studied  analytically and numerically
 in \cite{burb,burl}, and
here we will extend the previously
elaborated continuous-space
 and time
mean-field-type description
to the more complicated process under study.
 In this description
we will focus on the
evolution of the BP  mean displacement,
 which we denote as $R(t)$,
and mean occupation (or density) of
 the site $X$ at time $t$, $\rho(X,t)
= <\eta(X,t)>$, where brackets denote
 averages with respect to
 different realizations of
the stochastic process.

We start with the description of the dynamics of the boundary particle,
whose
 mean displacement is found to obey
the following exact equation

\[\tau \; \frac{d R(t)}{d t}  \; =
\; a \; p \; -  \; a \; q \; (1 \;  -
 \;  \rho_{1}), \quad  (6)\]
where $\rho_{1} = \rho(X =  R(t) - a,t)$, i.e. the mean
occupation of the site adjacent to the
position of the BP.

Turning now to the dynamics
of the film particles, we note that here
we have to consider separately
the evolution of $\rho(X,t)$ on
 sites $X$
of the interval [$0, R(t) - 2 a$] and at $X = R(t) - a$.
Particles
 which may be present at the
first interval  are all identical.
In contrast,
evolution of $\rho(X,t)$ at $X = R(t) - a$ is affected
by  the BP with its asymmetric
 transition rates.
For the first interval
any (forbidden) attempt
of any particle to hop onto the well already
occupied by another particle
 is quite
equivalent to the event when both simply
 interchange their positions,
 which means that
hard-core exclusion is not very important
 for
 the
evolution of  $\rho(X,t)$ on
[$0, R(t) - 2 a$]. Thus, as a
 reasonably good approximation we
suppose that
on this interval the density
$\rho(X,t)$ obeys a standard
diffusion equation

\[\frac{\partial \rho(X,t)}{\partial t} \; =
\; D \; \frac{\partial^{2}
\rho(X,t)}{\partial X^{2}}; \; D \; =
\; \frac{a^{2}}{4 \tau}, \quad (7)\]
while for the dynamics of $\rho(X,t)$ at $X = R(t) - a$ we will have

\[
a \frac{d \rho_{1}}{d t} \; =
\; - \; D \; \left. \frac{\partial \rho(X,t)}{\partial X}\right| _{ X = R(t) -
a} \;
- - \; \rho_{1} \; \frac{d R(t)}{d t},  \quad (8)
\]
in which the first term on the right-hand-side
accounts for the exchanges of identic
particles between
the sites $R(t) - 2 a$ and $R(t) - a$;
the second term describes the change in
the mean occupation of the site
 $X = R(t) - a$ due to the motion of the
 BP.  Here, the multiplier
 $d R(t)/d t$ determines the rate at which
the site adjacent to the BP becomes vacant
 due to the motion of the BP.
In turn,  the factor $\rho_{1}$, which is
the mean occupation of the site
adjacent to the  boundary particle,  accounts
 in a mean-field manner for the
following circumstance: Suppose that at
 time $t$ the BP is at $R(t)$ and
 the site $X =
R(t) - a$ is vacant, i.e. $\eta(X = R(t) - a,t) = 0$.
Then, if at the time moment $t + \tau$ the BP makes
a hop away from the MLE, it "creates" a
 vacancy at the previously occupied
 site and thus
$\eta(X = R(t) - a,t + \tau) = 0$ still
 equals zero. Therefore,
the occupation of this site is not
effectively changed in the case when prior
 to the BP hop the lhs
adjacent
site was vacant.
Conversely, if at time $t$ the lhs adjacent
 to the BP site is occupied, i.e.
$\eta(X = R(t) - a,t) =
1$,  and the BP hops
away of the MLE, one has that
 $\eta(X = R(t) - a, t + \tau) = 0$, i.e. is
 changed from one to zero.

Eqs.(6) to (8)  constitute a  complete, coupled system of
 dynamical equations
describing
the time evolution of the particle
 density $\rho(X,t)$ and the mean
 displacement
$R(t)$
of the boundary particle (extension of the film)  in
capillary rise geometries.

Consider now the appropriate extension of
these equations for the case of {\it circular drops} deposited on a horizontal
substrate.
In this case we will proceed essentially along the same lines, as in the case
of
capillary rise geometries. First, we will assume  that the MLE and the edge of
the film
are
ideal circular lines of radii $R_{mle}$ and  $R(t)$  respectively;
the fluctuations of the film edge are disregarded.
Further on, to specify the spatial positions
of the wells we introduce polar coordinates ($X,\phi$), where $X$ is the radial
coordinate and $\phi$ denotes the polar angle. Supposing next
that  density profiles
are radially symmetric, we will neglect the angular dependence of the density
profiles.
Turning now to the mathematical description of the problem, we notice that
under
such assumptions
 only Eq.(7) will be modified, while Eqs.(6) and (8) will remain the same,
provided that $X$
denotes now the radial variable.
Explicitly, Eq.(8)  will have the form

\[\frac{\partial \rho(X,t)}{\partial t} \; =
\; D \; [\frac{\partial^{2}}{\partial X^{2}} \; + \; \frac{1}{X} \;
\frac{\partial}{\partial X} ] \; \rho(X,t), \quad (9)\]
i.e.  the laplacian operator will be two-dimensional, in contrast to
the effectively one-dimensional diffusion operator which appears in the
capillary rise geometries.
Eq.(9) is to be solved subject to the boundary condition at the MLE,
$\rho(X = R_{mle},t) = \rho_{0}$, and Eq.(8). We note that, in fact, $R_{mle}$
is a slowly varying function of time (see Sec.1); its  time dependence,
however,
is uncomparably weaker than  an expected $\sqrt{t}$-law for growth
of the film radius (see Fig.4). Thus in the following we will assume that
$R_{mle}$
is constant.

\subsection{Solutions of dynamical equations
in case of capillary rise geometries.}

We will base our approach to the solution of coupled
nonlinear Eqs.(6) to (8)
on  $a$ $priori$ assumption that $R(t)$ actually
grows in time as $\sqrt{t}$ and that the
density profile
$\rho(X,t)$  attains  a stationary form in
terms of a scaled variable $\omega$, $\omega =
(X  - R_{mle})/(R(t) + a - R_{mle})$.
We note that, of course, the
solution so obtained
must be tested
for consistency with the initial assumption.
Consequently, such
 an approach
 will be self-consistent if we
succeed to show
that there exists
a finite, constant prefactor in the dependence
 $R(t) \sim \sqrt{t}$,
for which Eqs.(6) to (8) are compatible.

Rewriting Eq.(7) in terms of the defined above
scaled variable $\omega$ we have

\[D \frac{d^{2} \rho(\omega)}{d \omega^{2}} \; + \; \omega
\; (R(t) \; - \; R_{mle}) \; \frac{d R(t)}{d t}
 \;  \frac{d \rho(\omega)}{d \omega} \; = \; 0  \; \quad (10)\]
Let us denote

\[ A_{m} \; = \; \frac{1}{2 D} \; \frac{d (R(t) - R_{mle})^{2}}{d t} \quad
(11)\]
 In view of our assumption this parameter is expected to be a
time-independent constant, which (when found explicitly) will  define
the growth law of $R(t)$.

To find $A_{m}$ we will proceed as
follows: We
 notice first that since $d R(t)/d t \to 0$ when $t \rightarrow \infty$,
Eq.(6)
insures that $\rho_{1}$ rapidly, at rate $|d \rho_{1}/dt| \ll d R(t)/dt$,
approaches a
constant value $\tilde{\rho}_{1}$, $\tilde{\rho}_{1} = 1 - p/q$.
Then,  solving the differential Eq.(10)
subject to the boundary conditions $\rho(\omega = 0) = \rho_{0}$ and
$\rho(\omega = 1) =
\tilde{\rho}_{1}$ we find

\[\rho(\omega) \; = \; \rho_{0} \; +
\; (\tilde{\rho}_{1} \; - \; \rho_{0}) \; \frac{erf(\omega
\sqrt{A_{m}/2})}{erf(\sqrt{A_{m}/2})}, \quad (12)\]
where $erf(x)$ denotes the error function.  From Eq.(12) we then obtain

\[\left. \frac{d \rho(\omega)}{d \omega}\right| _{ \omega = 1} \; = \; - \;
\sqrt{\frac{2 A_{m}}{\pi}} \;
\frac{(\rho_{0} - \tilde{\rho}_{1}) \;
\exp(- A_{m}/2)}{erf(\sqrt{A_{m}/2})} \quad (13)\]
On the other hand,
 rewriting Eq.(8) in terms of
the scaled variable $\omega$ and neglecting
 transient terms, we will get

\[\left. \frac{d \rho(\omega)}{d \omega}\right| _{\omega = 1} \; = \; - \;
A_{m} \; \tilde{\rho}_{1} \quad (14)\]
On comparing the rhs of Eqs.(13) and (14) we arrive at a closed
with respect to $A_{m}$ equation,
 which defines its dependence on the given parameters
$E_{\downarrow}$ and $W_{\leftarrow}$,

\[\sqrt{\frac{\pi A_{m}}{2}} \; \exp(\frac{A_{m}}{2})
\; erf(\sqrt{\frac{A_{m}}{2}}) \;
 = \; \frac{1 \; -
 \; \exp( - \beta s)}{\exp( \beta W_{\leftarrow}) \; -
\; 1}, \quad (15)\]
in which $s$ denotes the difference
\[ s \; =  \; E_{\downarrow} \; - \; W_{\leftarrow} \quad (16)\]

A simple analysis shows that whenever the rhs of Eq.(15) is
positive, Eq.(15) has a single positive solution and thus $A_{m}$
is actually a well-defined positive constant.
Consequently,  we may claim that the mean
 displacement of
the boundary particle  (or, in other words,
the mean extension
 of the wetting film) obeys

\[R(t) \; = \; R_{mle} \; + \; \sqrt{2 \; A_{m}\;  D \; t}, \quad (17)\]
i.e. the form of Eq.(1) in which the "fitting" parameter
$\alpha$ (Sec.II) is equal to the product of the diffusion
coefficient and the parameter $A_{m}$.
Eq.(17) is the primary analytical result of our analysis
 and agrees with the
experimentally observed time dependence \cite{hesa,hesb,hesc,hesd,amcc}.

Now, we estimate analytically  the
dependence of $A_{m}$ on the pertinent parameters in
 the asymptotic
 limit
when $A_{m}$ is small or large. It follows from Eq.(15)
   that $A_{m}$ is small when the rhs of
Eq.(15) is small. It happens, namely, when
either the inequality $\beta E_{\downarrow} > \beta W_{\leftarrow}
\gg 1$  holds (what may be thought off as the case
 of liquids
with high cohesion energy and strong attraction
 to the substrate),
or when the parameter $s$
is sufficiently small, such that $\beta s \ll
\exp(\beta W_{\leftarrow}) - 1$.
When either of these
 inequalities
is fulfilled $A_{m}$ is given explicitly
by
\[A_{m} \; \approx \; \frac{(1
- - \exp(- \beta s))}{(\exp(\beta W_{\leftarrow}) - 1)}
  \quad (18)\]

We note now that growth of the film occurs
 as long as the
 parameter $s$, Eq.(16),
is positive, i.e. as long as $E_{\downarrow}$ exceeds the work $W_{\leftarrow}$
needed to transport a vacancy from the edge of the film to the macroscopic
liquid edge.
 Therefore,  the parameter $s$ is the key property which distinguishes
 whether the monolayer
will grow or not; thus it seems natural
 to define  $s$
as the $microscopic$ analogue of the spreading
 parameter $S$ -
the property  which rules  spreading of
 liquids at the
 macroscopic scales.
 Consequently, we will call $s$
 the
$microscopic$
spreading parameter.

Now,  $A_{m}$ may be large when the rhs of Eq.(15)
 is large,
 which happens  when $\beta W_{\leftarrow} \ll 1$ and $s$  is
sufficiently large.
In this case $A_{m}$ reads

\[A_{m} \; \approx \; - \; 2 \; ln(\beta W_{\leftarrow})
 \quad (19)\]
Behavior as in Eq.(19) may be realized
 experimentally in case of
liquids with low cohesion energy, which are
 volatile in two-dimensions,
but not volatile
in 3D. An example of such a liquid is
 squalane (see \cite{hesb} for details).

It may be worthwhile to remark that
 the behavior of the density profiles defined by Eq.(12) is
very different in the limits when $A_{m}$ is small or large. When $A_{m}$
is small $\rho(\omega,t)$ displays a linear dependence on $\omega$ (and thus on
$X$),
$\rho(\omega,t) \approx \rho_{0} +
(\tilde{\rho}_{1} - \rho_{0}) \omega$, while for the situations in which
$A_{m}$
is large
it shows much stronger variation with $\omega$. In neither case, however, one
can assume
the liquid monolayer in capillary rise geometries to be an incompressible
fluid.

Consider now the time evolution of the mass $M(t)$
of the film, defined as

\[ M(t) \; \approx \; \int^{R(t)}_{R_{mle}} dX \; \rho(X,t)  \quad (20)\]
Changing the variable of integration and making use of Eq.(12) we get

\[ M(t) \; \approx \; (R(t) - R_{mle}) \; \int^{1}_{0} d\omega \; \rho(\omega)
\; = \; \quad\]
\[ \; = \; (R(t) - R_{mle}) \; \exp(A_{m}/2) \; (1
- - \exp(- \beta W_{\leftarrow})),  \quad (21)\]
which shows that $M(t)$ also grows in proportion to $\sqrt{t}$, in
accord with experimental observations \cite{hesa,val}.
Eqs.(17) and (21) thus imply that mean density of particles in the film,
 $\overline{\rho}$, remains constant,
\[\overline{\rho} \;  =  \; \frac{M(t)}{(R(t) - R_{mle})} \; =
\; \exp(A_{m}/2) \; (1
- - \exp(- \beta W_{\leftarrow})) \quad (22)\]
In case of small $A_{m}$ the
 mean
 density is
close to unity, while for progressively large $A_{m}$
it tends to zero.
This behavior
is illustrated in  Fig.6, where we plot the functions
 $M(t)/\sqrt{t}$,
$(R(t) - R_{mle})/\sqrt{t}$ and $\overline{\rho}$ versus the
transition probability $q$, $q = 1/(1 + exp(- \beta W_{\leftarrow}))$.

We close this subsection with some comments concerning spreading kinetics
in situations in which $W_{\leftarrow} = 0$, i.e. when "liquid-vapour"
interface is absent.  This case is somewhat peculiar, since
the rhs of Eq.(15) diverges, which means that $A_{m}$ is no longer
a well-defined constant but rather is some increasing function of time.
Eq.(19) shows that $A_{m}$ diverges logarithmically when $W_{\leftarrow}$
tends to zero; thus one may expect that $A_{m}$ grows in proportion to $ln(t)$
when $W_{\leftarrow}$ is exactly equal to zero.  In \cite{burd} we have shown
that
it is actually so and $A_{m}$ displays a slow logarithmic growth
\[A_{m} \; \approx \; ln(\frac{4 \rho_{0}^{2} D t}{\pi a^{2}}) \quad (23)\]
at sufficiently large times.  This means, in fact, that the presence (or
absence)
of attractive liquid-liquid interactions (or of the SOSM interface in our
description)
does not affect significantly the spreading kinetics resulting only in slowly
varying in time prefactors.

\subsection{Solutions of dynamical equations for circular droplets on a
horizontal substrate}

Let us now examine  the growth law  of the precursor film for circular
droplets.
In this case the density profiles, defined by Eq.(9),
do not attain the stationary form $\rho(\omega)$; the gradient term
$X^{-1} \partial/\partial X$ in the laplacian operator does not allow to
represent
the complete time dependence of $\rho(X,t)$ in terms of
 the scaled variable $\omega$ only.
We thus shall seek for the solution of Eq.(9) of the form
$\rho(X,t) = \rho(\omega,t)$. For this, Eq.(9) reads

\[\frac{R_{mle}^{2}}{D \mu^{2}(t)} \; \frac{\partial \rho(\omega,t)}{\partial
t}
\; = \;
\frac{\partial^{2} \rho(\omega,t)}{\partial \omega^{2}} \; +
\; (\frac{1}{\omega + \mu(t)} \;
+ \; A_{m} \omega ) \;
\frac{\partial \rho(\omega,t)}{\partial \omega}, \quad (24)\]
in which we have denoted $\mu(t) = R_{mle}/(R(t) - a - R_{mle})$.  The solution
of Eq.(24) can be found
recursively, expanding $\rho(\omega,t)$
 in the inverse powers of
the diffusion coefficient
\[ \rho(\omega,t) \; = \; \sum_{n=0}^{\infty} D^{- n} \Psi_{n}(\omega,t) \quad
(25)\]
In doing so, we will obtain for the zeroth term

\[\Psi_{0}(\omega,t) \; = \; \rho_{0} \; + \; (\tilde{\rho}_{1} \; -
\; \rho_{0}) \; \int^{\omega}_{0}
 \frac{d \omega}{\omega + \mu(t)} \; \exp( -
\omega^{2} A_{m}/2) \;  \times \quad\]
\[ \times \;  \{\int^{1}_{0} \frac{d \omega}{\omega + \mu(t)} \; \exp( -
\omega^{2} A_{m}/2)\}^{-1}, \quad (26)\]
while higher-order terms will be defined  through

\[\frac{\partial^{2} \Psi_{n}(\omega,t)}{\partial \omega^{2}} \; +
\; (\frac{1}{\omega + \mu(t)} \;
+ \; A_{m} \omega ) \;
\frac{\partial \Psi_{n}(\omega,t)}{\partial \omega} \; = \;
\frac{R_{mle}^{2}}{\mu^{2}(t)} \; \frac{\partial \Psi_{n-1}(\omega,t)}{\partial
t} \quad (27)\]
Straightforward, but rather tedious analysis which will be presented elsewhere
\cite{burk},
shows however that only the zeroth term is relevant; the higher-order
$\Psi_{n}(\omega,t)$ define
only small at any $t$ and $\omega$ corrections and  $\rho(\omega,t) \approx
\Psi_{0}(\omega,t)$
occurs to be quite an accurate approximation.

Consider now behavior of $\Psi_{0}(\omega,t)$ defined by Eq.(26).  At
relatively
short times,
when $R(t)$ does not exceed significantly $R_{mle}$, the function $\mu(t) \gg
1$. Consequently,
in this time regime we may safely neglect  the variable
$\omega$ ($0 \leq \omega \leq 1$) compared to $\mu(t)$ in the function $(\omega
+ \mu(t))^{-1}$.
Then,  Eq.(26) simplifies to
the form of Eq.(12), which defines
the stationary density profiles in capillary rise geometries. This means, in
turn, that at early
stages of the film growth its radius obeys exactly the law in Eq.(17) with the
parameter $A_{m}$
defined by Eq.(15).  Of course, this result is not unexpected on physical
grounds - when
$R(t)$ is comparable to $R_{mle}$
effects of curvature can not be important.  Turning next to the opposite limit,
when $R(t) \gg
R_{mle}$ and thus $\mu(t) \ll 1$, we find from Eq.(26),

\[\left. \frac{d \rho(\omega,t)}{d \omega}\right| _{\omega = 1} \; \approx \; -
\;
(\rho_{0} - \tilde{\rho}_{1}) \; \exp( - A_{m}/2) \; \{\int^{1}_{0} \frac{d
\omega}{\omega +
\mu(t)} \; \exp( - \omega^{2} A_{m}/2)\}^{-1}  \quad (28)\]
Comparing Eqs.(28) and (14) we will obtain for the parameter
$A_{m}$:

\[\frac{1 - \exp(- \beta s)}{\exp(\beta W_{\leftarrow}) - 1} \; \approx \;
\exp(A_{m}/2) \; A_{m}
\; \int^{1}_{0} \frac{d \omega}{\omega + \mu(t)} \; \exp( -
\omega^{2} A_{m}/2) \quad (29)\]
Now, since $\mu(t) \to 0$ when $t$ progresses, the value of the integral in
Eq.(29) is dominated by
the lower limit, i.e. vicinity of $\omega = 0$. Neglecting then the exponent
$\exp( - \omega^{2}
A_{m}/2)$, which will contribute only to the second order in powers of $A_{m}$,
and integrating, we
get the following relation

\[\frac{1 - \exp(- \beta s)}{\exp(\beta W_{\leftarrow}) - 1} \; \approx \;
A_{m}
\;
ln(\frac{1}{\mu(t)}), \quad (30)\]
which yields for $t \gg R^{2}_{mle}/2 D$, ($R(t) \gg R_{mle}$),

\[A_{m} \; \approx \; \frac{1 - \exp(- \beta s)}{\exp(\beta W_{\leftarrow}) -
1}
\; \frac{2}{ln(2 D t/R^{2}_{mle})} \; \{ 1 \; + \;  \quad \]
\[ \;  + \; \frac{ln[ln(2 D t/R^{2}_{mle}) \; +
\; (\exp(\beta W_{\leftarrow}) - 1)/2 (1 - \exp(- \beta s))]}{ln(2 D
t/R^{2}_{mle})} \; +
\; ... \; \} \quad (31)\]
Eq.(31) thus shows that in this time limit the radius of the monolayer grows as
$R(t) \sim \sqrt{t/ln(t)}$, in accord with the
prediction of de Gennes and Cazabat \cite{pdgb}.  The prefactors in the growth
law in Eq.(31)
are, however,  different from these obtained in \cite{pdgb}.
The density profiles corresponding to this time
regime attain the form

\[\rho(X,t) \; \approx \; \rho_{0} \; + \; (\tilde{\rho}_{1} \; - \; \rho_{0})
\;
\frac{ln(X/R_{mle})}{ln(R(t)/R_{mle})}, \quad (32)\]
which shows a logarithmically slow variation with $X$. This means apparently
that in the regime
$R(t) \gg R_{mle}$ the liquid monolayer arising during spreading of a circular
droplet can be
approximately viewed as an  incompressible 2D fluid.

Finally, let us estimate the time evolution of the mass of particles in the
film
and of the
mean density. For the mass we obtain

\[M(t) \; = \; 2 \; \pi \; \int^{R(t)}_{R_{mle}} X \; dX \; \rho(X,t), \quad
(33)\]
where $\rho(X,t)$ is defined in the large-$t$ limit by Eq.(32). Substituting
Eq.(32) into the
Eq.(33) and integrating, we get

\[M(t) \; \approx \; \pi \; [(R^{2}(t) \; - \; R^{2}_{mle}) \; (\rho_{0} \; +
\;
\frac{\rho_{0} - \tilde{\rho}_{1}}{ 2 \; ln(R(t)/R_{mle})} \; + \; R^{2}(t) \;
(\tilde{\rho}_{1} -
\rho_{0}) ] \quad (34)\]
Consequently, the mean density of particles in the monolayer obeys

\[\overline{\rho} \; \approx \; \tilde{\rho}_{1} \; +
\; \frac{\rho_{0} - \tilde{\rho}_{1}}{ 2 \; ln(R(t)/R_{mle})}, \quad (35)\]
which means that in situations, in which the SOSM interface enclosing the
hard-sphere fluid is
present, the mean density of particles in the film tends to a constant value,
dependent on the
surface tension of the interface. In the absence of the interface (i.e. when
$W_{\leftarrow} = 0$),
the mean density  decreases in time in proportion to $\rho_{0}/ln(t)$.

\section{Summary and discussion}

To conclude, we have presented here both experimental
 and theoretical
analysis of kinetics
of thin liquid films spreading on solid interfaces.

In the experimental part we have described results of recent
 ellipsometric measurements
of the time dependent thickness profiles and of the growth
 rate of the first layer
on top of interface. Experiments, which were
carried out on different types of bare or grafted
solid substrates and were performed
with two different types of liquids - melts of light
polymers (PDMS)
and liquid of spherical rigid molecules (TK), have shown
 the extraction of a molecularly
thin precursor film, advancing  well ahead of the macroscopic
 liquid edge,  and also
displayed several other features observed
 in earlier works
\cite{hesa,hesb,hesc}. Namely,  the appearance of stepped,
"terraced" shapes of the liquid
droplet at the microscopic, molecular scales in case of the
 PDMS droplets and also, for both
types of liquids,  confirmed that the first layer on top of
solid grows in proportion to
$\sqrt{t}$.

We have proposed an analytical model in which the spreading monolayer
 is viewed
as a hard-sphere fluid enclosed by an effective "liquid-vapour"
interface which stabilizes
fluid and mimics, in a mean-field fashion, the presence of cohesive
liquid-liquid interactions.
The macroscopic drop in our description is considered as a reservoir
 of particles of an
infinite capacity, which feeds the film.  The model allows an
 analytical solution;
we have found explicit expressions describing the
 growth of the first layer
for both the case of capillary rise geometries and
circular droplets on a horizontal
solid, as well as determined the time evolution of
 the mass of particles in the film, the mean
density and dynamical density profiles. For the case
of capillary rise geometries we have shown
that $R(t)$ (the extension of the film above the
 macroscopic meniscus) grows in time as
$\sqrt{2 A_{m} D t}$, where $D$ is the "bare" diffusion
 coefficient describing random motion
of a particle on solid interface, $A_{m}$ is the parameter
 which is dependent on the
magnitude of
liquid-liquid and liquid-solid interactions. This parameter
 is determined implicitly, in form of
a transcendental equation; in limiting situations its explicit
forms are found.  Mass of
particles in the film is also shown to grow in proportion to
 $\sqrt{t}$, which means that
the mean density stays constant. We have also shown that local
 density of particles in the
film varies essentially with the altitude above the meniscus.
 Now, in case of circular
droplets on a horizontal substrate, our analytical findings
 are as follows: we predict that
at short times,
when the radius of the first layer is comparable to the radius
 of the macroscopic liquid edge,
the growth of the film occurs essentially like in the capillary
rise geometries. Within the
opposite limit, when $R(t)$ exceeds substantially $R_{mle}$, we
 have established that the
parameter $A_{m}$ decreases in time as $1/ln(t)$, which yields
$R(t) \sim \sqrt{t/ln(t)}$, in
accord with the theory of de Gennes and Cazabat \cite{pdgb}.
 We have also
shown that in this time regime the density profile is
 described by a logarithmically slow
function of the radial distance from the macroscopic
liquid edge, i.e. an assumption that
the monolayer can be viewed as an incompressible 2D fluid \cite{pdgb}
 may be physically
plausible when
$R(t) \gg R_{mle}$.
Explicit results for the mass of particles and for the mean density
are also presented.

Further on, our analytical results suggest that the physical
 mechanism responsible for
the $\sqrt{t}$-law for growth of monolayers is associated
 with the diffusive transport
of "vacancies" from the edge of the film to the macroscopic
liquid edge.  Arriving to the
MLE a vacancy perturbes  the equilibrium between the film and
 the bulk liquid; and then
is filled by a fluid particle from the reservoir.  We specified
the microscopic
spreading parameter "$s$",
which distinguishes whether the monolayer will grow or not.
In our picture, this parameter
is equal to the difference of the energy gained by filling
 a vacancy
at the macroscopic liquid edge by a fluid particle
and the work required to transport a vacancy from the edge
of the film to the MLE.  Growth of a monolayer does not take
 place if this parameter is
negative. In view of this we may comment that it is somewhat
 misleading to call the
$\sqrt{t}$ growth of monolayers as "diffusive"; here it
 describes the growth of the $mean$
displacement, i.e. spreading, which is exactly zero for
diffusive-type processes.  Spreading
of liquid monolayers is rather reminiscent, in view of
the physics involved, of the phenomena
of directional solidification or of melting, in which
 the spreading of  a "new" phase front
is controlled by the rate at which the particles of an
"old" phase diffuse away of it
 \cite{lang}.

Finally, we remark that the model discussed here
 has several evident shortcomings and
is to be improved in several directions.  First,
 assuming the reservoir to have an infinite capacity,
 we certainly
limit our model description
to only some intermediate time regime, which is, of
 course, quite
 extended in time but nonetheless does not cover all
stages
 of the liquid droplet spreading.
The third point of our model, which is  some sort of a
 mean-field type assumption,
 may be more bothering since it substitutes the collective
attractive interaction of all particles in the film
by some effective one, imposed on the
particles at the edge of the film only.
It is heuristically resolved
here by taking the effective surface tension  equal to
 the work required to
transport a vacancy from the edge of the film to the
 macroscopic liquid edge, but
surely a model involving explicitly the cohesive
interactions is required.
Let us discuss within the framework of the present
analytical approach
the final stage of spreading of a liquid droplet.
 In our approach,  allowing the reservoir to
 be exhausted, i.e. violating
 the boundary condition in Eq.(2)
at some moment of time, would yield immediate
 termination of the film growth;
the impenetrable "liquid-vapour" interface embracing
the molecules on solid interface will prevent
 further spreading
and the film will form a stable, circular liquid
patch of molecular thickness.  For monolayers this is the case
for 2D non-volatile liquids (PDMS with molecular mass above
$2000$).  For 2D volatile liquids
the
 molecular diffusion ultimately destroys such  structure \cite{hesa}.
Experiments performed with light PDMS molecules and
 squalane \cite{hesa} (see
also \cite{kon} and references therein) clearly show
 that a molecularly
thin liquid patch, appearing
after the central part of the droplet is emptied,  is not
stable and continues to spread indefinitely forming a
 two-dimensional gas.
Experimental data, described here in Fig.3,
show that the film ceases to grow when the reservoir
disappears and thus  seemingly
conforms to the prediction of our model. We remind,
 however, that the PDMS molecules
involved in this study
 were rather
long, containing more than one hundred monomers.
 Effects of the reduced diffusivity
or, possibly, entanglements,  may matter here and
 give rise to extremely slow spreading
inaccessible at
experimentally available time scales.

\begin{center}
\begin{Large}
Acknowledgments.
\end{Large}
\end{center}
S.F.B acknowledges the  support of ONR Grant N 00014-94-1-0647 and by
the University of Paris VI. G.O. acknowledges financial support from the
FNRS, Belgium.

\begin{center}
\begin{Large}
Figure Captions.
\end{Large}
\end{center}

Fig.1. Time evolution of the thickness profiles of a TK droplet.

Fig.2. Spreading of a TK droplet. Dynamical thickness profiles (top);
Growth of the radius $R(t)$ of the first layer vrs $\sqrt{t}$ (bottom).

Fig.3.  Spreading of a PDMS droplet. Dynamical thickness profiles (top);
Growth of the radius $R(t)$ of the first layer vrs $\sqrt{t}$ (bottom).

Fig.4.  Spreading of a PDMS droplet at high relative
humidity. Large droplet (top);
small droplet (bottom).

Fig.5. Schematic picture of  a liquid monolayer spreading on
 solid interface.
Wavy line depicts the effective "energy" landscape arising
 due to liquid-liquid and
liquid-solid interactions.

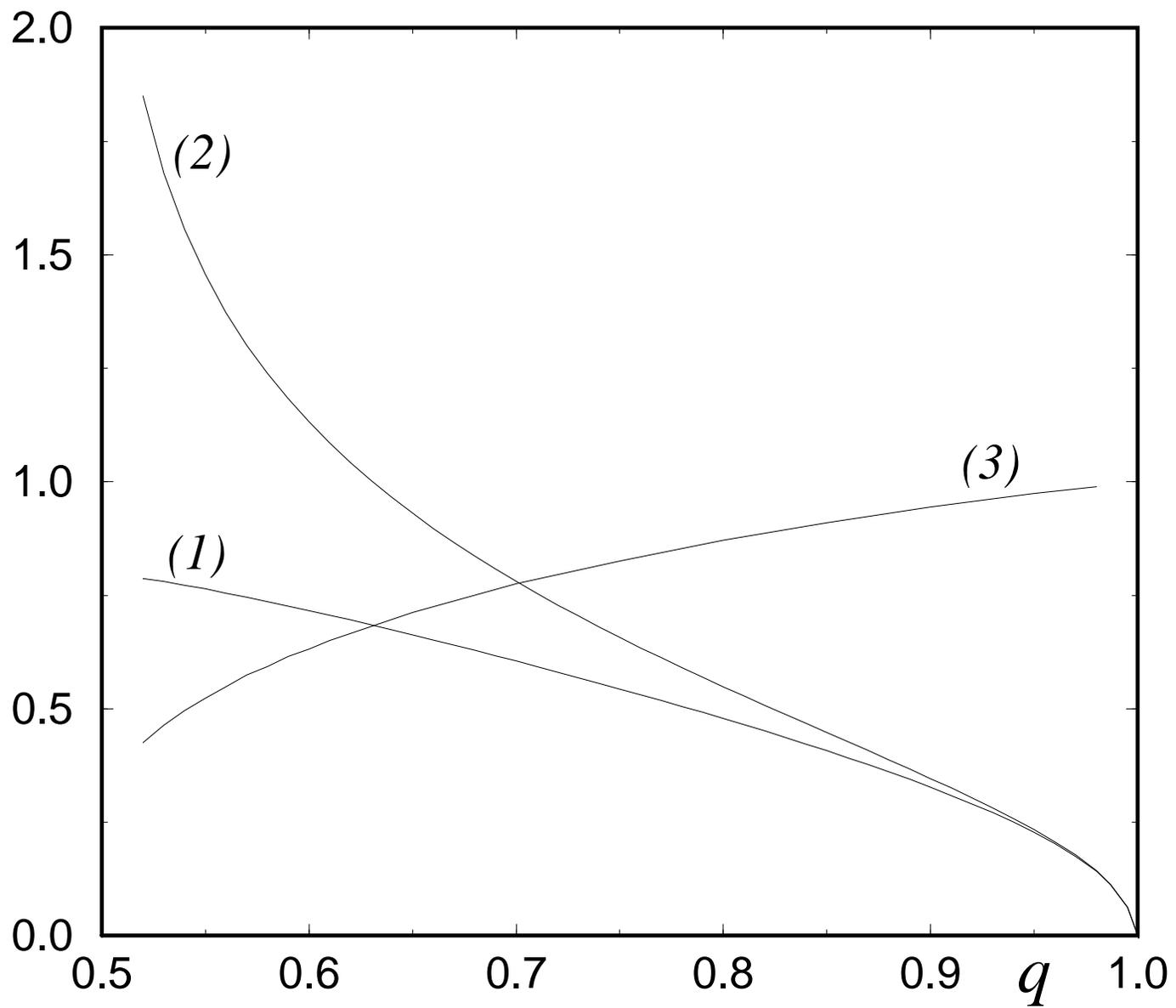
Fig.6.  Plot of the analytical dependences of the functions
$M(t)/\sqrt{t}$ (curve (1));
$(R(t) - R_{mle})/\sqrt{t}$ (curve (2)) and the  mean density
 $\overline{\rho}$ (curve (3))
versus the parameter $q = 1/(1 + \exp(\beta W_{\leftarrow}))$.

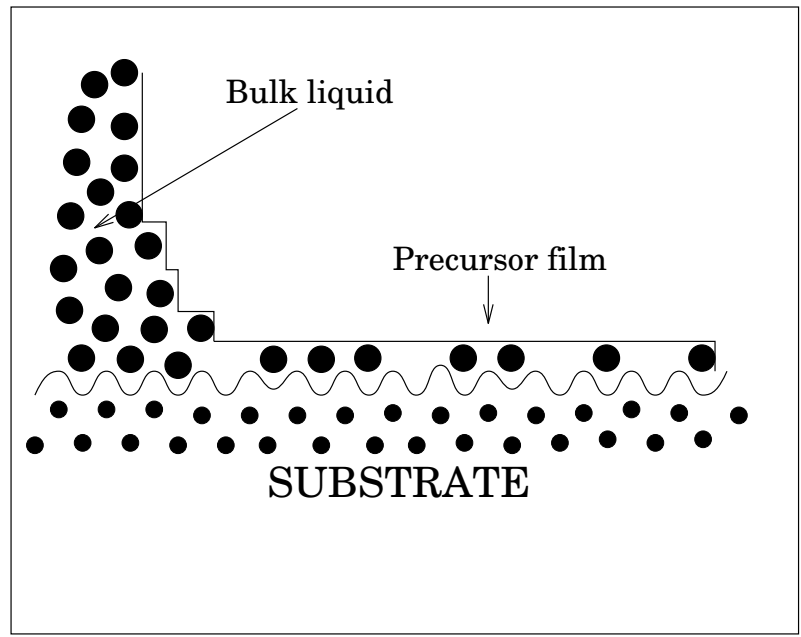
\end{document}